\documentclass[reprint,superscriptaddress,amsmath,amssymb,aps,prl,notitlepage,floatfix,nofootinbib,longbibliography,twocolumn]{revtex4-1}

\usepackage{changes}
\usepackage{tensor}     
\usepackage{graphicx}   
\usepackage[
colorlinks=true,        
citecolor=blue,         
linkcolor=blue,         
urlcolor=blue           
]{hyperref}             
\usepackage{bm}         
\usepackage{xcolor}     
\usepackage{lipsum}
\usepackage{color}      
\usepackage[utf8]{inputenc} 
\usepackage[section]{placeins} 
\usepackage{multirow}
\usepackage{orcidlink}
\newcommand{\nc}{\newcommand*} 

\nc{\al}{\alpha}
\nc{\s}{\sigma}
\nc{\kp}{\kappa}
\nc{\dt}{\delta}
\nc{\Dt}{\Delta}
\nc{\Ld}{\Lambda}
\nc{\p}{\partial}
\nc{\Gm}{\Gamma}
\nc{\om}{\omega}
\nc{\Om}{\Omega}
\nc{\rd}{\mathrm{d}}
\def\({\left(}
\def\){\right)}
\def\[{\left[}
\def\]{\right]}
\def\e{\begin{equation}}
\def\q{\end{equation}}
\def\m{\begin{eqnarray}}
\def\n{\end{eqnarray}}
\nc{\Eq}[1]{Eq.~\eqref{#1}}     
\nc{\Eqs}[1]{\eqref{#1}} 
\nc{\Fig}[1]{Fig.~\ref{#1}}     
\nc{\Table}[1]{Table~\ref{#1}}  
\nc{\Sec}[1]{Sec.~\ref{#1}}     
\nc{\Msun}{M_\odot}             
\nc{\fpbh}{f_{\mathrm{pbh}}}    
\nc{\fpbhn}{f_{\mathrm{pbh0}}}    
\nc{\mR}{\mathcal{R}} 
\nc{\seq}{\sigma_{\mathrm{eq}}}
\nc{\ogw}{\Omega_{\mathrm{GW}}}
\nc{\gpcyr}{\mathrm{Gpc}^{-3}\,\mathrm{yr}^{-1}}
\nc{\lvc}{LIGO/Virgo} 
\nc{\SNR}{\mathrm{SNR}} 
\nc{\mmin}{{m_{\mathrm{min}}}}
\nc{\mmax}{{m_{\mathrm{max}}}}
\nc{\Mmin}{{M_{\mathrm{min}}}}
\nc{\fmin}{{f_{\mathrm{min}}}}
\nc{\VT}{\mathrm{VT}}
\nc{\rhoGW}{\rho_{\mathrm{GW}}}
\nc{\vth}{\vec{\theta}}
\nc{\vd}{\vec{d}}
\nc{\vla}{\vec{\lambda}}
\nc{\Nobs}{N_{\mathrm{obs}}}
\nc{\av}[1]{\langle #1 \rangle} 
\nc{\km}{\mathrm{km}}
\nc{\Mpc}{\mathrm{Mpc}}
\nc{\Tobs}{T_{\mathrm{obs}}}
\nc{\Ntemp}{N_{\mathrm{temp}}}
\nc{\fyr}{f_{\mathrm{yr}}}
\nc{\addref}{[\textcolor{red}{add ref}] } 
\nc{\eg}{\textit{e.g.~}}
\nc{\app}{\approx}
\nc{\hf}{\frac{1}{2}}
\nc{\discuss}{\textcolor{red}{Add discussion here!}}
\nc{\red}[1]{\textcolor{red}{#1}}
\nc{\hp}{h_+} 
\nc{\hc}{h_{\times}} 
\nc{\Oh}{\hat{\Omega}}
\nc{\vx}{\vec{x}}
\nc{\mh}{\hat{m}}
\nc{\nh}{\hat{n}}
\nc{\zh}{\hat{z}}
\nc{\ph}{\hat{p}}
\nc{\A}[1]{\mathcal{A}_{#1}}
\nc{\Ogw}[1]{\Omega_{\mathrm{#1}}}
\nc{\bn}[1]{\dt{t}_{\text{#1}}}
\nc{\bC}[1]{{C}_{\text{#1}}}
\nc{\NTOA}{N_{\text{TOA}}}
\nc{\Nmode}{{N_{\text{mode}}}}
\nc{\ARN}{A_{\rm{RN}}}
\nc{\gRN}{\gamma_{\rm{RN}}}
\nc{\bS}{{\Sigma}}
\nc{\br}{{r}}
\nc{\bN}{{R}}
\nc{\Agw}{A_\mathrm{GWB}}
\nc{\UCP}{\mathrm{UCP}}
\nc{\TT}{\mathrm{TT}}
\nc{\ST}{\mathrm{ST}}
\nc{\SL}{\mathrm{SL}}
\nc{\VL}{\mathrm{VL}}
\nc{\lpr}{l^{\prime}}
\nc{\mpr}{m^{\prime}}
\nc{\Cth}{\mathcal{C}_{\mathrm{th}}}
\newcommand{\E}{\mathbb{E}}
\newcommand{\Var}{\operatorname{Var}}
\newcommand{\Cov}{\operatorname{Cov}}

\newcommand{\dd}{\mathrm{d}}
\newcommand{\CN}{\mathcal{CN}}
\newcommand{\Id}{\mathbb{I}}
\newcommand{\Lap}{\mathcal{L}}
\newcommand{\norm}[1]{\left\lVert#1\right\rVert}

\nc{\BFST}{$107 \pm 7$}
\begin{document}
	

\title{Joint Angular-Power Distributions of Interfering Nanohertz Gravitational Waves}

\author{Yan-Chen Bi\orcidlink{0000-0002-9346-8715}}
\affiliation{Institute of Theoretical Physics, Chinese Academy of Sciences, Beijing 100190, China}
\affiliation{School of Physical Sciences, University of Chinese Academy of Sciences, No. 19A Yuquan Road, Beijing 100049, China}

\author{Yu-Mei Wu\orcidlink{0000-0002-9247-5155}}
\email{Corresponding author: wuyumei@yzu.edu.cn} 
\affiliation{Center for Gravitation and Cosmology, College of Physical Science and Technology, Yangzhou University, Yangzhou, 225009, China}

\author{Qing-Guo Huang\orcidlink{0000-0003-1584-345X}}
\email{Corresponding author: huangqg@itp.ac.cn}
\affiliation{Institute of Theoretical Physics, Chinese Academy of Sciences, Beijing 100190, China}
\affiliation{School of Physical Sciences, 
    University of Chinese Academy of Sciences, 
    No. 19A Yuquan Road, Beijing 100049, China}
\affiliation{School of Fundamental Physics and Mathematical Sciences, Hangzhou Institute for Advanced Study, UCAS, Hangzhou 310024, China}

\begin{abstract}
Finite populations of supermassive black-hole binaries
produce fluctuations in both gravitational-wave power
and pulsar timing correlations.
We characterize the realization-dependent,
orientation-averaged Earth-term angular correlation
through its frequency-resolved Legendre coefficients.
A hybrid model of bright sources and a Gaussian
faint background extends the conditional Gamma
distribution in the many-weak-source limit to
non-central $\chi^2$ statistics, yielding correlated
multipoles after population marginalization.
Our analytical variance decomposition and SMBHB
population simulations reveal a population-dependent
transition from interference-dominated variance
at low frequencies to population-power-dominated
variance at high frequencies.
In the many-weak-source regime, a multivariate
lognormal approximation captures the main features
of the joint distribution using only the first
two log-cumulants.
These results provide statistical benchmarks
for fluctuations in both the overall amplitude
and angular structure of the realized correlation.
\end{abstract}

\maketitle


\emph{Introduction.} \label{sec:introduction}
Pulsar timing arrays (PTAs) probe nanohertz gravitational waves (GWs) through correlated variations in the pulse arrival times of Galactic millisecond pulsars~\cite{1990ApJ...361..300F}. Current PTA analyses commonly model the gravitational-wave background (GWB) as a stationary, isotropic, unpolarized Gaussian process, whose ensemble spatial correlations follow the Hellings--Downs (HD) curve. This Gaussian approximation is well motivated when each frequency bin contains sufficiently many weak sources with independent random phases and no dominant contribution. However, an astrophysical GWB generated by a finite population of supermassive black hole binaries (SMBHBs) need not satisfy these conditions.

SMBHBs are discrete, slowly evolving GW sources in the PTA band. The discreteness of their population leads to realization-to-realization fluctuations in the total GW power, with rare nearby or massive binaries producing a heavy high-power tail that becomes more prominent at high frequencies. 
At low frequencies, a single Fourier bin may contain $\gtrsim{\cal O}(10^3)$ binaries~\cite{Allen:2022dzg}. Their radiation-reaction timescales, $t_c\sim{\cal O}(10^6)\,{\rm yr}$, far exceed typical PTA observing baselines, $T_{\rm obs}\sim{\cal O}(10)\,{\rm yr}$~\cite{Corbin:2010kt}, allowing multiple sources to remain within the same Fourier bin throughout the observation. Their complex amplitudes then add coherently, producing interference-induced fluctuations in the angular correlation pattern. Consequently, a single astrophysical realization need not reproduce either the ensemble-averaged GW spectrum or the HD correlation.

These effects have been studied from several complementary perspectives. The variance and cosmic variance of the HD correlation have been derived for Gaussian and discrete-source ensembles~\cite{Allen:2022dzg,Allen:2022ksj,Bernardo:2022xzl},
while interference between discrete sources has been shown to generate frequency-dependent multipolar departures from the HD curve\cite{Wu:2024xkp,Wu:2025xqh,Bi:2026jeu}. More recently, non-Gaussian statistics of the strain power, timing residuals, and higher-order correlators have been investigated~\cite{Xue:2024qtx,Lamb:2025niq,Domcke:2025esw, Ali-Haimoud:2026sbk,Raidal:2026ezm,Kuntz:2026usl,Hisamatsu:2026qvd,Cecchini:2026cru,Falxa:2026mhh}. 
Together, these studies highlight that a realistic astrophysical GWB is not fully characterized by its ensemble-averaged two-point function. A more complete statistical description should therefore characterize the probability distribution of the realized angular covariance itself, accounting for both interference-induced fluctuations in the angular correlation pattern and population-induced variations in the total GW power.




In this Letter, we develop a statistical description of the realized angular correlation using frequency-resolved Legendre multipoles of the pulsar-averaged Earth-term correlation. We construct a hybrid model of discrete bright sources and a Gaussian faint background, extending the conditional Gamma distribution in the many-weak-source limit to non-central $\chi^2$ statistics when bright sources become important.
Marginalizing over source-population fluctuations yields the joint multipole distributions and correlations between different angular modes. We further quantify the contributions of interference and total-power fluctuations to the multipole variance, and derive a compact multivariate lognormal approximation in the many-weak-source regime. We test these predictions against Monte Carlo SMBHB populations.

\emph{Frequency-resolved angular spectrum.} 
\label{sec:onefreq}
Consider a frequency bin $p$ centered at $f_p$, with complex source amplitudes $s_j^p = \sqrt{2/3}\mathcal{A}_j e^{i\Phi_j} \operatorname{sinc}[\pi T(f_j-f_p)]$, where $\mathcal{A}_j = \sqrt{2/5} A_j$ is the effective circularly polarized amplitude and $A_j$ is the physical face-on amplitude. In the narrow-bin approximation, we replace the sinc response by a delta function, $\operatorname{sinc}[\pi T(f_j-f_p)] \propto \delta(f_j-f_p)$, neglecting spectral leakage and inter-bin correlations. We henceforth consider a single frequency bin and suppress the index $p$.

Let $\tilde{z}_a(t)$ be the complex Earth-term redshift, expanded as $\tilde{z}_a(t) = \sum_{\ell m}z_{\ell m} Y^*_{\ell m}(\hat{\boldsymbol p}_a)$. For a given GW realization, averaging pulsar pairs over all orientations at fixed angular separation
$\gamma_{ab}$ defines the angular correlation~\cite{Allen:2024bnk}
\e
C(\gamma_{ab}) =
\langle \tilde{z}_a \tilde{z}^*_b \rangle_{\gamma_{ab}}
= \sum_{\ell=2}^{\infty} S_\ell P_\ell(\cos\gamma_{ab}).
\q
Using the spin-weighted harmonic representation of the GW field, the Legendre coefficients are given by
\e
S_\ell = c_\ell^{\rm HD} \frac{4\pi}{n_\ell}
\sum_{m=-\ell}^{\ell} \vert {}_2a_{\ell m} \vert^2,
\label{eq:sl}
\q
where $n_\ell = 2\ell+1$, $c_\ell^{\rm HD} = 3 n_\ell/[2(\ell-1)\ell(\ell+1)(\ell+2)]$, and ${}_2a_{\ell m} = (-1)^\ell \sum_j s_j {}_2Y_{\ell m}(\hat{\boldsymbol\Omega}_j)$. The HD coefficients satisfy $\sum_\ell c_\ell^{\rm HD} P_\ell(\cos\gamma_{ab}) = \Gamma_{\rm HD}(\gamma_{ab})$, with $\Gamma_{\rm HD}(0) = 1/2$. 

Substituting the source expansion yields
\begin{align}
S_\ell &= c_\ell^{\rm HD} H_2 + I_\ell,
\quad \text{with} \quad H_n = \sum_j \vert s_j \vert^n,
\label{eq:split}
\end{align}
where
\begin{align}
I_\ell &= \frac{8 \pi c_\ell^{\rm HD}}{n_\ell}
\Re \sum_m \sum_{j<k} s_j s_k^*
{}_2Y_{\ell m}(\hat{\boldsymbol\Omega}_j)
{}_2Y_{\ell m}^*(\hat{\boldsymbol\Omega}_k).
\end{align}
The common total-power term $c_\ell^{\rm HD}H_2$ preserves the HD multipole proportions, while the interference term $I_\ell$ introduces multipole-dependent deviations, with $C(\gamma_{ab})=H_2\Gamma_{\rm HD}(\gamma_{ab})$ in the absence of interference.
Thus, fluctuations in $S_\ell$ arise from both population-power fluctuations and wave interference.

To isolate relative interference-induced angular fluctuations, we define the power-normalized multipole $Q_\ell\equiv S_\ell/(c_\ell^{\rm HD}H_2) = 1+I_\ell/(c_\ell^{\rm HD}H_2)$. 
For many weak sources with isotropic directions and independent uniform phases, the proper-complex Gaussian approximation gives $Q_\ell\sim\operatorname{Gamma} (n_\ell,n_\ell)$, or $2n_\ell Q_\ell\sim\chi^2_{2n_\ell}$~\cite{Allen:2024uqs}, where the Gamma parameters denote shape and rate. This approximation can fail when a few bright SMBHBs dominate.

To account for bright sources, we construct a hybrid model in which bright sources are retained explicitly and the faint population is approximated by an isotropic, unpolarized, proper-complex Gaussian field. The harmonic mode vector is
\begin{equation}
\bm B_\ell =
\sqrt{\frac{4\pi}{n_\ell}}
({}_2a_{\ell \ell}, {}_2a_{\ell \ell-1}, \cdots,
{}_2a_{\ell -\ell})^T
= \bm b_\ell^B + \bm u_\ell,
\label{eq:hybrid}
\end{equation}
where $\bm u_\ell\sim\CN(0,H_u\Id/n_\ell)$, $H_u$ is the faint-source power, and
$S_\ell=c_\ell^{\rm HD}\norm{\bm B_\ell}^2$.
The Gaussian vectors $\bm u_\ell$ are independent across $\ell$ at fixed $H_u$.
For a fixed bright-source realization, let $W_\ell^B=\norm{\bm b_\ell^B}^2$.
The conditional Laplace transform of $S_\ell$, $\Lap_\ell(t) = \E[e^{-tS_\ell}\mid\bm b_\ell^B,H_u]$, follows from Gaussian integration:
\begin{equation}
\Lap_\ell(t)=
\frac{\exp\left[
-\dfrac{t c_\ell^{\rm HD}W_\ell^B}
{1+t c_\ell^{\rm HD}H_u/n_\ell}
\right]}
{(1+t c_\ell^{\rm HD}H_u/n_\ell)^{n_\ell}}.
\label{eq:non-centralLT}
\end{equation}

\begin{figure}[tpb]
\centering
\includegraphics[width=.45\textwidth]{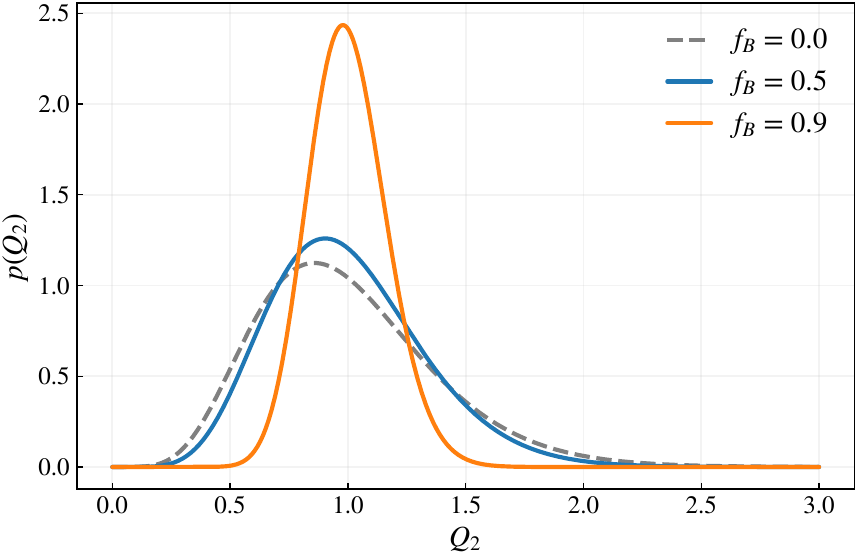}
\caption{Conditional distributions of the power-normalized
quadrupole $Q_2$ for different bright-source fractions
$f_B$. The gray dashed curve represents the
central Gamma limit ($f_B=0$), while the blue
and orange solid curves show the non-central
$\chi^2$ distributions at $f_B=0.5$ and $0.9$,
respectively.
Increasing $f_B$ narrows the distribution
while preserving its unit mean, indicating
suppressed interference-induced angular fluctuations.
\label{fig:ql_demo}}
\end{figure}

For multiple bright sources, $W_\ell^B$ includes their mutual interference and generally depends on $\ell$. We focus on the single-bright-source limit, motivated by individual-source searches ~\cite{Goncharov:2026cdh} and the possible dominance of one binary in high-frequency bins~\cite{Ellis:2023dgf}.
In this limit, $W_\ell^B=H_B$ for all $\ell$. Defining the total power $H_2=H_u+H_B$ and bright-source fraction $f_B=H_B/H_2$, Eq.~(\ref{eq:non-centralLT}) yields a scaled non-central $\chi^2$ distribution for $Q_\ell$,
\begin{equation}
Q_\ell\mid f_B \sim
\frac{1-f_B}{2n_\ell}
\chi'^2_{2n_\ell}
\left(\frac{2n_\ell f_B}{1-f_B}\right).
\label{eq:fractiondist}
\end{equation}
At fixed total power, increasing $f_B$ suppresses interference-induced angular scatter without changing the mean, as $\E[Q_\ell\mid f_B]=1$ and $\Var[Q_\ell\mid f_B]=(1-f_B^2)/n_\ell$.
The distribution reduces to the central Gamma law at $f_B=0$ and approaches $\delta(Q_\ell-1)$ as $f_B\to1$, as shown in \Fig{fig:ql_demo}.


For a Poisson source population, the total GW power $H_2$ follows a compound-Poisson distribution~\cite{Ali-Haimoud:2026sbk,Sato-Polito:2024lew,Raidal:2026ezm,Yu:2026rba}. 
Given the joint population distribution $p_H(H_2,f_B)$, marginalizing over both $H_2$ and $f_B$ yields
\begin{equation}
p(S_\ell)=
\int \dd H_2\int_0^1\dd f_B\,
p_H(H_2,f_B)\,
p(S_\ell\mid H_2,f_B).
\label{eq:B_marginal}
\end{equation}
where the conditional distributions follow from Eq.~(\ref{eq:fractiondist}) through $S_\ell=c_\ell^{\rm HD}H_2Q_\ell$. This expression applies to the positive-power component. Allowing both $H_u$ and $H_B$ to fluctuate, the law of total covariance gives
\begin{align}
\Cov(S_\ell,S_{\ell'})
= c_\ell^{\rm HD}c_{\ell'}^{\rm HD}
\bigg[
&\frac{\delta_{\ell\ell'}}{n_\ell}
\E(H_u^2+2H_uH_B)
+\Var(H_2)
\bigg].
\label{eq:randombright}
\end{align}
The first term describes interference-induced angular fluctuations at fixed source powers, while the second arises from fluctuations in the total power $H_2$ and correlates different multipoles.
In the single-source limit, angular fluctuations vanish, but fluctuations in $H_2$ can still contribute to the covariance. Equation~(\ref{eq:randombright}) recovers known finite-source covariances under the corresponding assumptions~\cite{Allen:2024mtn} (see Supplemental Material).

\begin{figure}[tpb]
\centering
\includegraphics[width=0.45\textwidth]{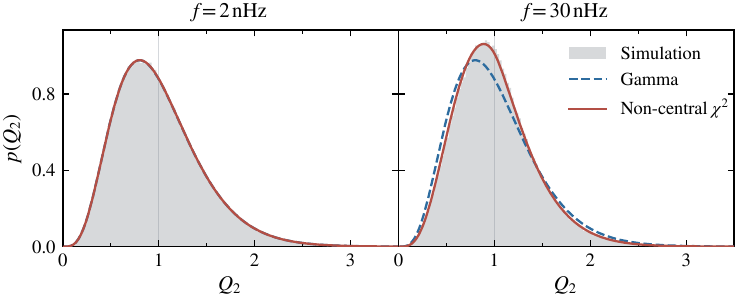}
\caption{Conditional distributions of the power-normalized quadrupole $Q_2$ at low and high frequencies, obtained by randomizing source phases and sky directions at fixed source amplitudes. Gray histograms show simulations, while blue dashed and red solid curves represent the central Gamma and conditional non-central $\chi^2$ predictions, respectively. Both models agree at low frequencies, but the noncentral $\chi^2$ model performs better at high frequencies.
\label{fig:ql}}
\end{figure}

We test these predictions using Monte Carlo realizations of SMBHB populations based on semi-analytic galaxy and black-hole evolution prescriptions~\cite{Chen:2018znx,Bi:2023tib,NANOGrav:2023hfp} (see Supplemental Material). At $f=2\,\mathrm{nHz}$ and $30\,\mathrm{nHz}$, we fix the source amplitudes and randomly sample their phases and sky directions, keeping $H_2$ and $f_B$ unchanged. \Fig{fig:ql} compares the resulting distributions of $Q_2$ with the central Gamma and conditional non-central $\chi^2$ predictions. Both models agree with simulations at low frequencies, whereas the non-central $\chi^2$ model provides a better description at high frequencies, where bright sources become important.

\begin{figure*}[htpb!]
\centering
\includegraphics[width=.9\textwidth]{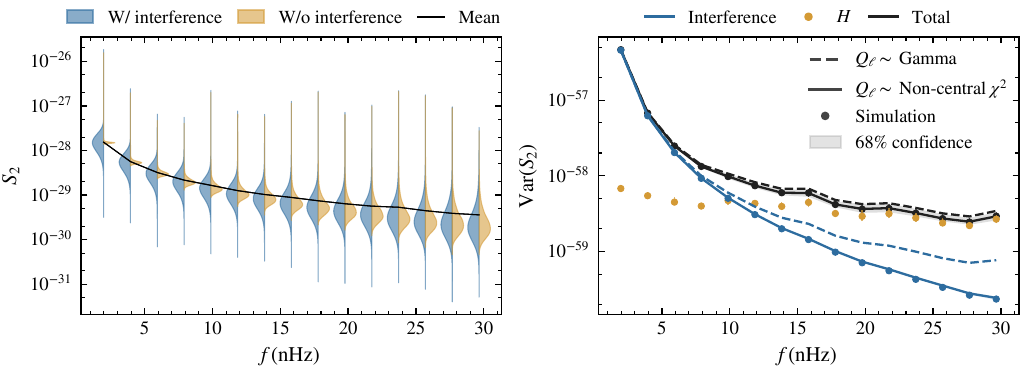}
\caption{\textbf{Left panels}: Distributions of the angular power $S_2$ at different frequencies, with (blue) and without (orange) wave interference. Interference broadens the distributions without changing their means, with a diminishing effect at higher frequencies. \textbf{Right panels}: Total variance of $S_2$ and its contributions from interference and population-power fluctuations as a function of frequency. Dots show simulations, while dashed and solid curves show predictions based on the Gamma and non-central $\chi^2$ models for $Q_2$, respectively. The non-central $\chi^2$ model agrees with simulations across the frequency range, whereas the Gamma approximation deviates at high frequencies. Interference dominates the variance at low frequencies, while population-power fluctuations dominate at high frequencies. \label{fig:sl_freq}}
\end{figure*}

We then allow the source population to vary between realizations, introducing fluctuations in both $H_2$ and $f_B$. Marginalizing over their joint distribution $p_H(H_2,f_B)$ yields the distributions of $S_2$, incorporating both wave interference and total-power fluctuations. 
The left panels of \Fig{fig:sl_freq} compare the distributions with and without interference, the latter corresponding to $S_2=c_2^{\rm HD}H_2$. Interference broadens the distributions without changing their means, with a diminishing effect at higher frequencies. The right panels show the total variance of $S_2$ and its decomposition into interference and population-power contributions.
Interference dominates at low frequencies, whereas population-power fluctuations dominate at high frequencies as rare bright binaries become more influential. This crossover depends on the population model. 
The non-central $\chi^2$ model reproduces the simulated variances across the frequency range, while the Gamma approximation deviates at high frequencies.

\textit{Lognormal approximation.} 
\label{sec:lognormal}
At low frequencies, where numerous weak sources contribute without a single dominant source, the harmonic modes are approximately Gaussian. The resulting conditional Gamma distribution allows a compact description of the population-marginalized Legendre coefficients in logarithmic space.

Writing $S_\ell=c_\ell^{\rm HD}H_2Q_\ell$, with $Q_\ell\sim\operatorname{Gamma}(n_\ell,n_\ell)$ independent of $H_2$, the cumulant-generating function of $\ln S_\ell$ is
\begin{align}
K_\ell(t)&\equiv\ln\E[S_\ell^t]
\label{eq:logB_cgf}\\
&=t\ln(c_\ell^{\rm HD}\overline H_2)
+K_{\ln h}(t)+\ln\frac{\Gamma(n_\ell+t)}{\Gamma(n_\ell)}
-t\ln n_\ell.\nonumber
\end{align}
where $\overline H_2\equiv\E[H_2]$ is the ensemble-averaged total GW power, $h=H_2/\overline H_2$ is the ensemble-normalized power, and $K_{\ln h}(t)=\ln\E[h^t]$.
Differentiating Eq.~(\ref{eq:logB_cgf}) at $t=0$ gives
\begin{align}
\kappa_1(\ln S_\ell)
&=\ln(c_\ell^{\rm HD}\overline H_2)
+\kappa_1(\ln h)+\psi(n_\ell)-\ln n_\ell,
\nonumber\\
\kappa_n(\ln S_\ell)
&=\kappa_n(\ln h)+\psi_{n-1}(n_\ell),
\quad n\geq2,
\label{eq:logcumulants}
\end{align}
where $\psi$ and $\psi_n$ denote the digamma and polygamma functions. The higher-order log-cumulants separate the two sources of fluctuations: $\kappa_n(\ln h)$ describes fluctuations in the total source power, while $\psi_{n-1}(n_\ell)$ describes angular realization fluctuations at fixed power.

Retaining the first two log-cumulants yields
\begin{equation}
\ln S_\ell\simeq{\cal N}(\mu_\ell,\sigma_\ell^2), \quad
\mu_\ell=\kappa_1(\ln S_\ell), \quad
\sigma_\ell^2=\kappa_2(\ln S_\ell).
\label{eq:lognormal}
\end{equation}
These parameters match the logarithmic rather than ordinary moments of $S_\ell$. At large $\ell$, the angular log-cumulants satisfy
$\psi_1(n_\ell)\sim n_\ell^{-1}$,
$\psi_2(n_\ell)\sim -n_\ell^{-2}$, and
$\psi_3(n_\ell)\sim 2n_\ell^{-3}$, so the standardized skewness and excess kurtosis vanish as ${\cal O}(n_\ell^{-1/2})$ and ${\cal O}(n_\ell^{-1})$.
Thus, the angular contribution approaches Gaussianity, while the full lognormal approximation also depends on the higher log-cumulants of $H_2$.

At fixed $H_2$, different multipoles are independent in the conditional Gamma model, but share the same random power after population averaging. Their leading joint approximation is therefore multivariate lognormal,
\begin{equation}
(\ln S_2,\ln S_3,\ldots)
\simeq{\cal N}(\boldsymbol\mu,\boldsymbol\Sigma),\ 
\Sigma_{\ell\ell'}=
\Var[\ln h]+\delta_{\ell\ell'}\psi_1(n_\ell),
\label{eq:multivariate_lognormal}
\end{equation}
where $\mu_\ell=\kappa_1(\ln S_\ell)$.
The common covariance term reflects population-power fluctuations, while the diagonal term describes independent interference-induced angular fluctuations. The accuracy of the lognormal approximation ultimately depends on the higher log-cumulants of $H_2$.

\begin{figure}[htpb!]
\centering
\includegraphics[width=.45\textwidth]{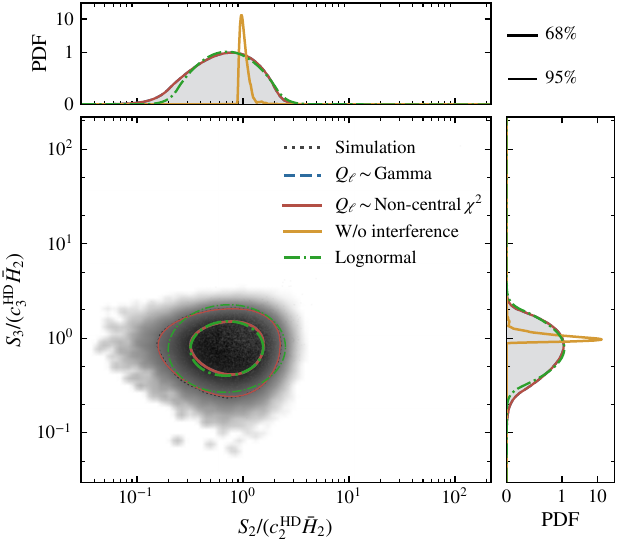}
\caption{Joint distribution of the ensemble-normalized Legendre coefficients $S_{2}/{(c_2^{\rm HD} \bar{H}_2)}$ and $S_{3}/{(c_3^{\rm HD} \bar{H}_2)}$ at $f=2 {\rm nHz}$. The central panel shows the joint PDF, with the corresponding marginal distributions shown in the top and right panels. Contours enclose the 68\% and 95\% probability regions. Predictions based on Gamma (blue dashed lines) and non-central $\chi^2$ (red solid lines) models for $Q_\ell$ both agree well with the simulations in this low-frequency regime. The lognormal approximation (green dot-dashed) is similarly accurate, whereas neglecting wave interference (orange solid lines) substantially underestimates the realization-to-realization scatter. \label{fig:sl_corner}}
\end{figure}

\Fig{fig:sl_corner} compares the simulated joint
distribution of ensemble-normalized Legendre coefficients $S_2/(c_2^{\rm HD}\bar H_2)$ and $S_3/(c_3^{\rm HD}\bar H_2)$ at $f=2\,\mathrm{nHz}$ with predictions obtained by marginalizing the conditional Gamma and non-central $\chi^2$ models over the source-power distribution, as well as the multivariate lognormal approximation. 
While less accurate than the full non-central $\chi^2$ model, the lognormal approximation captures the main features of the joint distribution, including correlations between multipoles, using only the first two log-cumulants. It thus provides a compact alternative to explicit population marginalization. Neglecting interference, by contrast, yields substantially narrower distributions.

\textit{Conclusions.} \label{sec:conclusions}
PTAs probe the GW background through correlated variations in pulsar timing signals. 
In this Letter, we have characterized the realization-dependent Earth-term angular correlation through its frequency-resolved Legendre coefficients $S_\ell$.
These coefficients separate into a common total-power term $c_\ell^{\rm HD}H_2$ and a wave-interference term $I_\ell$, describing fluctuations in the overall amplitude and angular structure of the correlation.

In the low-frequency regime dominated by numerous weak sources, the power-normalized multipoles $Q_\ell=S_\ell/(c_\ell^{\rm HD}H_2)$ follow a conditional Gamma distribution with shape and rate $n_\ell=2\ell+1$. When bright sources become important, a hybrid model yields non-central $\chi^2$ statistics that capture their effects on angular fluctuations. Marginalizing over source-population fluctuations gives the joint distributions of $S_\ell$ and correlations between different multipoles. Our SMBHB population simulations reveal a transition from interference-dominated variance at low frequencies to population-power-dominated variance at high frequencies.
In the many-weak-source regime, we further derive a multivariate lognormal approximation that captures the main features of the joint distributions using only the first two log-cumulants.

These results provide statistical benchmarks for the realized angular correlation, complementing the ensemble-averaged GW spectrum and Hellings--Downs correlation.
Extending this framework to fixed, noisy pulsar arrays requires incorporating their response and measurement uncertainties, which we leave for future work.

\begin{acknowledgments}
QGH is supported by the National Natural Science Foundation of China (Grant No.~12547110,~12475065, 12447101). YMW is supported by the National Natural Science Foundation of China (Grant No.~12505086).
	
\end{acknowledgments}

\bibliography{./ref}


\onecolumngrid
\appendix
\newpage

\begin{center}
{\LARGE\textbf{SUPPLEMENTAL MATERIAL}}
\end{center}

\begin{center}
{\Large
Yan-Chen Bi,$^{1,2}$ Yu-Mei Wu,$^{3}$ and Qing-Guo Huang$^{1,2,4}$}
\end{center}

\begin{center}
{\large\textit{$^1$Institute of Theoretical Physics, Chinese Academy of Sciences (CAS), Beijing 100190, China}}
\end{center}
\begin{center}
{\large\textit{$^2$University of Chinese Academy of Sciences (UCAS), Beijing 100049, China}}
\end{center}
\begin{center}
{\large\textit{$^3$Center for Gravitation and Cosmology, College of Physical Science and Technology, Yangzhou University, Yangzhou, 225009, China}}
\end{center}
\begin{center}
{\large\textit{$^4$School of Fundamental Physics and Mathematical Sciences, Hangzhou Institute for Advanced Study (HIAS), University of Chinese Academy of Sciences (UCAS), Hangzhou 310024, China}}
\end{center}


\section{SMBHB population model}\label{appb}
We summarize the population prescription used for the Monte Carlo calculation. The ensemble-mean squared characteristic strain, $\bar h^2_c(f)$, from SMBHBs is
\e
\bar h^2_c(f) = \int d M d q d z \frac{d^4N}{dz M dq d\ln f_r} {h^2(f_r)} ,
\q
where $M$, $q$ and $z$ are the primary black-hole mass, mass ratio, and redshift, and $f_r=(1+z)f$ is the rest-frame GW frequency. The sky- and polarization-averaged GW spectral strain of a single circular binary is
\e
h(f_r) = \frac{8}{\sqrt{10}} \frac{(G \mathcal{M})^{5/3}}{c^4 d_c} (\pi f_r)^{2/3},
\q
with chirp mass $\mathcal{M}$ and comoving distance $d_c$. The number of binaries per logarithmic rest-frame frequency interval is related to the galaxy differential merger rate by
\e
\frac{d^4N}{dz dM dq d\ln f_r} = \frac{d^3n_G}{dz dM_G dq_G} \frac{dM_G}{dM} \frac{dq_G}{dq} \tau(f_r) 4\pi c (1+z) d_c^2.
\q
Here, $c$ is the speed of light, $M_G$ and $q_G$ are the primary galaxy mass and the mass ratio between the paired galaxy. The adopted black-hole-bulge relation gives $q = q_G^{\alpha_*}$ with $\alpha_*$ a model parameter. For binaries evolving only through GW emission, the binary hardening timescale is given by
\begin{equation}
\tau(f_r) = f_r \frac{dt_r}{d f_r} = \frac{5}{96} \left(\frac{G \mathcal{M}}{c^3}\right)^{-5/3} (\pi f_r)^{-8/3}.
\end{equation}
Following \cite{Sesana:2008xk,Chen:2018znx}, the differential galaxy merger rate is
\e
\frac{d^3n_G}{dz dM_G dq_G} = \frac{\Phi(M_G,z^{\prime})}{M_G \ln 10} \frac{\mathcal{F}(M_G,z^{\prime},q_G)}{\tau(M_G,z^{\prime},q_G)} \left\vert\frac{dt_r}{dz^{\prime}}\right\vert,
\label{galaxy_model}
\q
where $z^{\prime}$ is the redshift at which the galaxy pair forms. It is related to the binary-merger redshift $z$ by
\e
\int_{z}^{z^{\prime}} \left\vert\frac{dt}{d\tilde{z}}\right\vert d\tilde{z} = \tau(M_G,z^{\prime},q_G) ,
\q
where the relationship between time and redshift assumes a flat $\Lambda$CDM Universe.
The merger timescale is parameterized as \cite{Snyder:2017timescale}
\begin{equation}
\tau(M_G,z^{\prime},q_G) = \tau_0 \(\frac{M_G}{b M_{G0}} \)^{\alpha_{\tau}} \( 1+z^{\prime} \)^{\beta_{\tau}} q_G^{\gamma_{\tau}},
\label{timescale}
\end{equation}
where $\tau_0$ represents merger time normalization with unit of $\rm{Gyr}$. Meanwhile, $bM_{G0} = 0.4 \times 10^{11} \Msun/h_0$ is an reference mass.

The galaxy stellar mass function (GSMF) is represented by a single Schechter form \citep{Mortlock:2014GSMF, Conselice:2016GSMF} 
\e
\Phi(M_G,z^{\prime}) = 10^{\Phi(z^{\prime})} (\ln 10) \(\frac{M_G}{M_{G0}}\)^{1 + \alpha(z^{\prime})} \exp{\(-\frac{M_G}{M_{G0}}\)},
\label{GSMF}
\q
where $\Phi(z^{\prime}) = \Phi_0 + z^{\prime} \Phi_I$, and $\alpha(z^{\prime}) = \alpha_0 + z^{\prime} \alpha_I$. 
The differential pair fraction is \cite{Chen:2018znx}
\e
\mathcal{F}(M_G,z^{\prime},q_G) = f_0^{\prime} \( \frac{M_G}{a M_{G0} } \)^{\alpha_f} \(1+z^{\prime} \)^{\beta_{f}} q_G^{\gamma_f},
\label{pair}
\q
where $aM_{G0} = 10^{11} \Msun$ is an reference mass \cite{Chen:2018znx}. 
The galaxy mass $M_{G}$ is mapped to the primary black-hole mass $M_{\mathrm{BH}}$ through bulge relation \cite{Sesana:2013wja, Kormendy:2013dxa}. We use $M_{\rm bulge} = f_{*,\text{bulge}} M_G$ and
\begin{equation}
M_{\mathrm{BH}} = \mathcal{N}\left\{ M_*\(\frac{M_{\rm bulge}}{10^{11} \Msun} \)^{\alpha_{*}} , \epsilon \right\},
\label{MMbulge}
\end{equation}
where $M_*$, $\alpha_*$, and $\epsilon$ are the normalization, slope, and logarithmic scatter. $\mathcal{N}\{x,y\}$ a log normal distribution with mean $x$ and standard deviation $y$. The fiducial parameters, calibrated to the NANOGrav 15-year data set, are listed in Table. \ref{Fiducial_values}.

\begin{table*}[tbp]
    \centering
    \setlength{\tabcolsep}{12pt}
    \caption{Fiducial values of parameters used in the analyses.}
    \label{Fiducial_values}
    \begin{tabular}{c c c c}
        \hline\hline
        \textbf{Model Component} & \textbf{Parameters} & \textbf{Description} & \textbf{Fiducial values} \\
        \hline
        \multirow{ 5}{*}{GSMF~\cite{Chen:2018znx}}    & $\Phi_0$  & GSMF norm & -2.78 \\
                    & $\Phi_I$  & GSMF norm redshift evolution & -0.27 \\
                    & $M_{G0}$   & GSMF scaling mass & +11.3 \\
                    & $\alpha_0$  & GSMF mass slope & -1.24 \\
                    & $\alpha_I$  & GSMF mass slope redshift evolution & -0.03 \\
        \hline
        \multirow{ 4}{*}{Pair fraction~\cite{Chen:2018znx}}    & $f^{\prime}_0$  & pair fraction norm  & 0.04 \\
                    & $\alpha_f$  & pair fraction mass slope  & +0.07 \\
                    & $\beta_{f}$  & pair fraction redshift slope  & 1.20 \\
                    & $\gamma_f$  & pair fraction mass ratio slope & 0.002 \\
        \hline
        \multirow{ 4}{*}{Merger timescale~\cite{Chen:2018znx}}    & $\tau_0$  & merger time norm  & 0.67 \\
                    & $\alpha_{\tau}$  & merger time mass slope  & -0.12 \\
                    & $\beta_{\tau}$  & merger time redshift slope & -1.69 \\
                    & $\gamma_{\tau}$  & merger time mass ratio slope & +0.007 \\
        \hline
        \multirow{3}{*}{$M_\text{bulge}-M_{\text{BH}}$~\cite{Kormendy:2013dxa}} & $\alpha_*$   & $M_\text{bulge}-M_{\text{BH}}$ relation slope   & +1.02 \\
                    & $M_{*}$   & $M_\text{bulge}-M_{\text{BH}}$ relation norm & 8.38 \\
                    & $f_{*,\text{bulge}}$   & $M_{G}-M_{\text{bulge}}$ relation ratio & 0.615 \\
                    & $\epsilon$  & $M_\text{bulge}-M_{\text{BH}}$ relation scatter  & 0.37 \\
        \hline
    \end{tabular}
\end{table*}

\section{Monte Carlo Realizations}\label{appc}
The SMBHB populations are generated with \textit{holodeck}~\cite{NANOGrav:2023hfp}, which follows semi-analytic prescriptions prescriptions above and the calibration of \cite{Chen:2018znx,Bi:2023tib}. These populations produce an expectation value for the number of binaries as a function of their parameters. To generate individual realizations of the Universe, we sample from a Poisson distribution around this expected value
\e
h^2_c(f_p) \!=\!\!\! \!\sum_{M,q,z,f}\!\!\mathcal{P} \left( \Delta M \Delta q \Delta z \Delta \ln f \frac{d^4N}{dz dM dq d\ln f_r} \right) \frac{h^2(f_p)}{\Delta \ln f} ,
\q
where $\mathcal{P}$ represents the Poisson sampling process. This sampling produces one realization of the binary population. Source directions and initial phases are then drawn independently and isotropically.



\section{Legendre representation of the GWB in PTAs} 
\label{sec:ng}
The GW contribution to pulsar-timing data can be described either by the redshift $Z_a(t)$ of pulsar $a$, or equivalently by the timing residual $r_a(t)=\int^t dt'\,Z_a(t')$. Throughout this work, we use the complex analytic redshift, with the physical redshift given by its real part. For a superposition of discrete sources,
\e
Z_a(t) = \sum_j \mathcal A_j e^{i(2\pi f_j t+\Phi_j)} R_a(\hat{\boldsymbol\Omega}_j,f_j),
\label{eq:redshift_time}
\q
where $f_j$, $\Phi_j$, and $\hat{\boldsymbol\Omega}_j$ denote the observed GW frequency, initial phase, and propagation direction of GW source $j$, respectively. Each binary is taken to be monochromatic over the PTA observing baseline, $h_j = \mathcal A_j \exp[{i(2\pi f_j t+\Phi_j)}]$, with polarization-averaged amplitude 
\begin{equation}
\mathcal A_j = \sqrt{\frac{32}{5}} \frac{(G\mathcal M_{z,j})^{5/3}(\pi f_j)^{2/3}}{c^4d_{L,j}}, 
\end{equation}
where $\mathcal M_{z,j}=(1+z_j)\mathcal M_j$ and $d_{L,j}$ are the redshifted chirp mass and luminosity distance of the source. The full pulsar response can be written as
\m
R_a(\hat{\Omega}_j, {f}_j) = \[1 - \mathrm{e}^{-i2\pi {f}_j L_a(1 + \hat{\boldsymbol\Omega}_j \cdot \hat{\boldsymbol p}_a) / c } \] F_a(\hat{\boldsymbol\Omega}_j) ,
\n
with the antenna pattern function, $F_a(\hat{\boldsymbol \Omega}_j)$ is given in terms of complex polarization tensor, $\varepsilon_{\alpha \beta}^{+,\times}$, as
\m
F_a(\hat{\boldsymbol\Omega}_j)=\frac{\hat{\boldsymbol p}_a^{a}\hat{\boldsymbol p}_a^{b}}{2(1+ \hat{{\boldsymbol\Omega}}_j \cdot \hat{\boldsymbol p}_a)} (\varepsilon_{\alpha \beta}^{+} - i \varepsilon_{\alpha \beta}^{\times}).
\n
where $L_a$ and $\hat{\boldsymbol p}_a$ are the distance and sky direction of pulsar $a$. The polarization tensors have the standard normalization $\epsilon^A_{\alpha \beta}\epsilon^{B,\alpha \beta}=2\delta_{AB}$. 
For an observing baseline $T$, we define the Fourier coefficient
\m
\tilde{Z}_a^p = \frac{1}{T} \int_{-T/2}^{T/2} Z_a(t) \mathrm{e}^{-i 2 \pi f_p t} dt = \sum_j \mathcal A_j e^{i\Phi_j} \operatorname{sinc} \left[\pi T(f_j-f_p) \right] R_a(\hat{\boldsymbol\Omega}_j,f_j), 
\n
where $f_p = p/T$, and $p \in \{1, 2, \cdots, N_f\}$ denotes Fourier modes with $N_f$ is the number of frequency bins. In the basis of spherical harmonics, the Earth term of redshift $\tilde{z}_a$ could be further expressed as
\begin{equation}
\tilde{z}_a^p = 4\pi \sum_{\ell m} \sqrt{\frac{c_\ell^{\mathrm{HD}}}{n_\ell}} {}_2 a_{\ell m}^p Y_{\ell m}^*(\hat{\boldsymbol p}_a),
\end{equation}
where $n_\ell = 2 \ell + 1$, and ${}_2a_{lm}^p = (-1)^\ell \sum_j s_{j}^p {}_2Y_{lm}(\hat{\boldsymbol\Omega}_j)$ with $s_{j}^p = \sqrt{2/3} \mathcal A_j e^{i\Phi_j} \operatorname{sinc}\!\left[\pi T(f_j-f_p)\right]$. In the narrow-bin approximation, we can safely replace the sinc response by an idealized frequency selection, $\operatorname{sinc}[\pi T(f_j-f_p)] \propto \delta(f_j-f_p)$, neglecting spectral leakage and correlations between frequency bins. We henceforth consider a single bin and suppress the index $p$. Noted that the factor $\sqrt{2/3}$ keeps the HD normalization. The HD multipole coefficients are
\begin{equation}
c_\ell^{\rm HD} = \left\{
\begin{aligned}
    &0 , &\text{for} & \ \ell=0,1 \\
    &\frac{3(2\ell+1)}{2} \frac{(\ell-2)!}{(\ell+2)!} , &\text{for} & \ \ell \geq 2 .
\end{aligned}
\right.
\end{equation}
Averaging over rotations of a pulsar pair at fixed angular separation $\gamma_{ab}$ gives
\begin{equation}
\left\langle Y_{\ell m}(\hat{\boldsymbol p}_a)Y_{\ell'm'}^*(\hat{\boldsymbol p}_b)\right\rangle_{\gamma_{ab}} = \frac{\delta_{\ell\ell'}\delta_{mm'}}{4\pi}P_\ell(\cos\gamma).
\end{equation}
Consequently, the orientation-averaged Earth-term angular correlation reads
\begin{equation}
C(\gamma_{ab}) = \Re\langle\tilde z_a\tilde z_b^{*}\rangle_{\gamma_{ab}} = \sum_{\ell\ge2}S_{\ell}P_\ell(\cos\gamma), 
\end{equation}
where the angular-correlation multipoles are
\begin{equation}
S_{\ell}=\frac1{4\pi}\sum_m|z_{\ell m}|^2 \ge 0 .
\label{eq:S}
\end{equation}
A useful normalized vector and equivalent spin notation are
\begin{equation}
b_{\ell m}=\frac{z_{\ell m}}{\sqrt{4\pi c_\ell^{\rm HD}}} = \sqrt{\frac{4\pi}{n_\ell}}\,{}_2a_{\ell m}.
\label{eq:normalized}
\end{equation}
The Legendre coefficients can then be written as
\begin{equation}
S_{\ell} = c_\ell^{\rm HD} \norm{\bm B_{\ell}}^2 = c_\ell^{\rm HD} \frac{4\pi}{n_\ell} \sum_m|{}_2a_{\ell m}|^2 = c_\ell^{\rm HD} \frac{4\pi}{n_\ell} \sum_m \left| \sum_j s_{j} {}_2Y_{lm}(\hat{\boldsymbol\Omega}_j)\right|^2.
\label{eq:angular-power}
\end{equation}
If wave interference is neglected, the coherent sum is replaced by an incoherent sum of individual source powers,
\begin{equation}
S_{\ell}^{\rm non-intf} = c_\ell^{\rm HD} \frac{4\pi}{n_\ell} \sum_m \sum_j |s_{j} {}_2Y_{lm}(\hat{\boldsymbol\Omega}_j)|^2 = c_\ell^{\rm HD} \sum_j |s_{j} |^2. 
\label{eq:nointer}
\end{equation}
Thus, the distinction between the interference and non-interference cases is whether source amplitudes are summed coherently before taking the power or their powers are summed incoherently.


\section{Angular-power distribution of an interfering GWB}
We define the source-power moments within a frequency bin as $H_{n} \equiv \sum_j |s_{j} |^n$. 
For isotropically distributed GW sources with independent random phases,
\begin{equation}
\int \frac{\dd \Phi}{2\pi} \int \frac{\dd \hat{\boldsymbol\Omega}}{4\pi} \left({}_2a_{\ell m} {}_2a_{\ell' m'}^{*} \right) = \frac{H_{2}}{4\pi} \delta_{\ell\ell'} \delta_{mm'}.
\label{eq:alm_covariance}
\end{equation}
If the effective number of sources, $N_{\rm eff}$, contributing to the frequency bin is large and no single source dominates, $\max_j {|s_{j }|^2}/{H_{2 }} \ll 1$, the central-limit theorem implies Gaussian approximation
\begin{equation}
{}_2a_{\ell m} \,\big|\, H_{2 } \simeq {\cal CN}\!\left(0,\frac{H_{2 }}{4\pi} \right).
\label{eq:alm_gaussian}
\end{equation}
Importantly, this is a Gaussian approximation for the angular coefficients at fixed total power. It does not imply that $H_2$ itself is Gaussian, the single-source power distribution is heavy tailed and can generate strongly non-Gaussian population-to-population fluctuations in $H_2$. 
For the properly normalized complex Gaussian distribution in \Eq{eq:alm_gaussian}, it follows 
\e
\frac{4\pi}{H_{2 }} \left|{}_2a_{\ell m} \right|^2 \sim \mathrm{Gamma}(1,1).
\q
where the Gamma distribution is parameterized by shape and rate. Summing over the $n_\ell=2\ell+1$ independent $m$ modes therefore gives the power-normalized multipole $Q_\ell = {S_{\ell }}/{(c_\ell^{\rm HD} H_{2})}$ as 
\begin{equation}
Q_\ell \vert H_{2 } \sim {\rm Gamma}(n_\ell,n_\ell).  
\end{equation}
Equivalently,
\begin{align}
p(S_{\ell }|H_{2 }) = \frac{n_\ell^{n_\ell}} {\Gamma(n_\ell)} \frac{S_{\ell }^{n_\ell-1}} {(c_\ell^{\rm HD} H_{2 })^{n_\ell}} \exp\!\left[-\frac{n_\ell S_{\ell }} {c_\ell^{\rm HD} H_{2 }}
\right],
\label{eq:B_conditional}
\end{align}
with $\Gamma(n_\ell)$ denotes the gamma function. This distribution also implies that the $S_{\ell }>0$ should be fulfilled. Thus, only positive values on $S_{\ell }$ are physically valid. 
Moreover, the corresponding skewness and (excess) kurtosis are
\begin{align}
\operatorname{Skew}(Q_{\ell}|H_{2 }) = \frac{2}{\sqrt{n_\ell}}, \\
\operatorname{Kurt}(Q_{\ell}|H_{2 }) = \frac{6}{{n_\ell}}.
\end{align}

\section{Bright-source conditional distribution}
Realistic SMBHB populations span a broad range of amplitudes, and one or a few binaries can contribute an appreciable fraction of the GW power in a frequency bin. We therefore separate the coherent field into an explicit bright-source contribution and an unresolved faint component, the latter being approximated as a proper complex Gaussian field. For a single explicit bright source,
${}_2a_{lm}  = (-1)^\ell \left[s_{\rm cw}  {}_2Y_{\ell m}(\hat{\boldsymbol\Omega}_{\rm cw}) + \sum_{j} s_j  {}_2Y_{\ell m}(\hat{\boldsymbol\Omega}_j)\right]$. The former one denotes contributions from bright source, while the latter still denotes unresolvable GWBs. The Legendre coefficients can correspondingly be decomposed as
\e
S_{\ell } = S_{\ell }^{\rm B} + S_{\ell }^{\rm u} + I_{\ell }.
\q
The first two terms denote the bright-source and unresolved-background contributions, while the wave-interference term $I_\ell$ contains their interference. Thus, a bright source does not simply add power to the background. It interferes coherently with the unresolved field.

At fixed complete bright-source catalog $\bm b^B_\ell$ and unresolved-background power $H_u$, write $\bm B_\ell = \bm b^B_\ell + \bm u_\ell$, with $\bm u_\ell \sim \CN(0, H_u\Id/n_\ell)$, and define $W^B_\ell=\norm{\bm b^B_\ell}^2$. The conditional proper-complex Gaussian density is
\begin{equation}
p(\bm{B}_\ell \mid \bm b^B_\ell, H_u) = (\pi H_u/n_\ell)^{-n_\ell} \exp[-\norm{\bm B_\ell - \bm b^B_\ell}^2n_\ell/H_u].
\end{equation}
Utilizing the polynomial variation as
\begin{equation}
\frac{\norm{\bm B-\bm b}^2}{v}+t c_\ell^{\rm HD}\norm{\bm B}^2 = \left(v^{-1}+t c_\ell^{\rm HD}\right) \norm{\bm B-\frac{\bm b}{1+t c_\ell^{\rm HD} v}}^2 + \frac{t c_\ell^{\rm HD}\norm{\bm b}^2}{1+t c_\ell^{\rm HD} v}.
\end{equation}
The integral of the shifted complex Gaussian gives Laplace transform 
\begin{equation}
\mathcal L_\ell(t \mid \bm b^B_\ell, H_u) = (1+t c_\ell^{\rm HD} H_u/n_\ell)^{-n_\ell} \exp\left[-\frac{t c_\ell^{\rm HD} W^B_\ell}{1+t c_\ell^{\rm HD} H_u/n_\ell}\right].
\label{eq:hybridLT}
\end{equation}
Each real or imaginary part of $u_{\ell m}$ has variance $H_u/(2n_\ell)$. Consequently,
\begin{equation}
\frac{2S_\ell n_\ell }{ c_\ell^{\rm HD} H_u} \mid \bm b^B_\ell, H_u \sim \chi'^2_{2n_\ell}\left(\frac{2 n_\ell W^B_\ell}{H_u}\right).
\end{equation}
Thus, the Gamma distribution is recovered as the zero-non-centrality limit of the more general non-central family. 
Differentiating the log transform gives conditional cumulants,
\begin{equation}
\kappa_n(S_\ell\mid\bm b^B_\ell,H_u) = (n-1)!( c_\ell^{\rm HD})^n \left[n_\ell v_\ell^n+nW^B_\ell v_\ell^{n-1}\right].
\label{eq:hybridcumulants}
\end{equation}
In particular, the mean is $\E(S_\ell|\bm b^B_\ell,H_u) = c_\ell^{\rm HD}(H_u + W^B_\ell)$ and the variance is $\Var(S_\ell|\bm b^B_\ell,H_u) = ( c_\ell^{\rm HD})^2(H_u^2 + 2H_u W^B_\ell)/n_\ell$. 
Defining $\theta_\ell= c_\ell^{\rm HD} H_u/n_\ell$ and $\eta_\ell=W^B_\ell n_\ell/H_u$, the conditional density may also be written in the standard non-central Gamma expansion,
\begin{equation}
p(S_\ell\mid\bm b^B_\ell,H_u) = e^{-\eta_\ell}\sum_{k=0}^{\infty}\frac{\eta_\ell^k}{k!} \frac{S_\ell^{n_\ell+k-1}e^{-S_\ell/\theta_\ell}} {\Gamma(n_\ell+k)\theta_\ell^{n_\ell+k}}.
\end{equation}
The Poisson index $k$ in this expansion is an auxiliary mathematical variable.

At fixed complete bright-source catalog, the unresolved Gaussian components are independent between multipoles. The population-averaged joint Laplace transform is therefore
\begin{equation}
\mathcal L(\bm t) = \E_{\bm b^B_\ell, H_u} \left[ \prod_\ell \mathcal L_\ell (t_\ell|\bm b^B_\ell, H_u) \right].
\label{eq:catalogmix}
\end{equation}
Importantly, we need to emphasis that conditioning only on the individual bright-source powers is insufficient when several explicit sources are retained. Because their contributions add coherently in $\bm b_\ell^B$, the quantity $W_\ell^B=\|\bm b_\ell^B\|^2$ also depends on their sky directions and relative phases. These variables must therefore be marginalized jointly over the complete bright-source catalog. Independently marginalizing the multipole densities would erase this shared catalog dependence.

\begin{figure*}[htpb!]
\centering
\includegraphics[width=1.\textwidth]{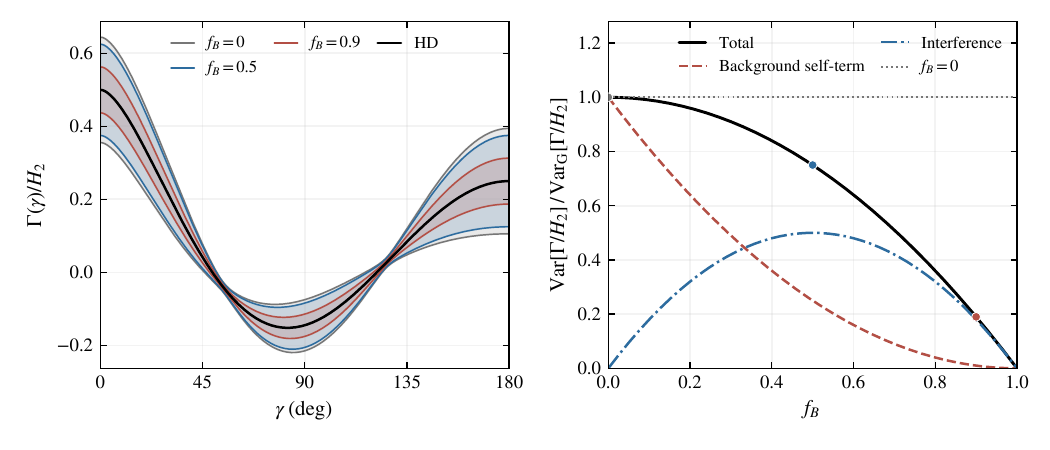}
\caption{Demonstration of cosmic variance for different values of $f_B$. Effect of a bright source on the realization-dependent angular correlation.
\textbf{Left panel}: Normalized correlation $\Gamma(\gamma)/H_2$ for representative bright-source fractions $f_B$ values, compared with the HD expectation.
\textbf{Right panel}: Variance of $\Gamma/H_2$ (black solid line), normalized to the Gaussian-background value (grey dotted line), together with its background self-term (red dashed line) and wave-interference term (blue dash-doted line). 
Increasing $f_B$ suppresses the total angular realization variance, which vanishes in the deterministic single-source limit.
\label{fig:cv_demo}}
\end{figure*}

For one source, $W^B_\ell = H_{B}$ for every $\ell$. Therefore, their physical meaning is more clear and practical. If $H_u$ is fixed, Eq.~\eqref{eq:catalogmix} reduces to
\begin{equation}
\mathcal L(\bm t)=\prod_\ell(1+t_\ell c_\ell^{\rm HD} v_\ell)^{-n_\ell} \mathcal L_{H_{B}}\left(\sum_\ell\frac{t_\ell c_\ell^{\rm HD}}{1+t_\ell c_\ell^{\rm HD} v_\ell}\right).
\end{equation}
Defining the total power $H_2 = H_u + H_{B}$ and bright-source fraction $f_B = H_{B}/H_2$, the power-normalized multipole gives
\begin{equation}
Q_\ell \mid f_B \sim \frac{1-f_B}{2n_\ell} \chi'^2_{2n_\ell}\left(\frac{2n_\ell f_B}{1-f_B}\right) ,
\label{eq:fraction}
\end{equation}
Its mean and variance are $\E[Q_\ell] = 1$, and $ \Var[Q_\ell] = {(1-f_B^2)}/{n_\ell}$. 
An alternative parameter is $\eta_B=H_{B}/H_u$, with $f_B=\eta_B/(1+\eta_B)$. The limit $f_B\to0$ recovers the Gamma distribution, whereas at the limit $f_B\to1$, Eq.~\eqref{eq:fraction} approaches $\delta(Q-1)$ corresponding to the deterministic single-source limit (CW-like) after pulsar averaging. \Fig{fig:cv_demo} illustrates how the angular variance evolves with the bright-source fraction $f_B$, showing its progressive suppression as the signal becomes increasingly dominated by a single coherent source.
Moreover, the higher moments are 
\begin{align}
\operatorname{Skew}(Q_{\ell}|f_B) = \frac{2}{\sqrt{n_\ell}} \frac{\sqrt{(1 - f_B)}(1 + 2f_B)}{(1 + f_B)^{3/2}}, \\
\operatorname{Kurt}(Q_{\ell}|f_B) = \frac{6}{{n_\ell}} \frac{(1 - f_B)(1 + 3f_B)}{(1 + f_B)^{2}}.
\end{align}

If both $H_u$ and $H_B$ fluctuate across population realizations and may be statistically dependent, the law of total covariance gives
\begin{equation}
\Cov(S_\ell,S_{\ell'})= c_\ell^{\rm HD} c_{\ell '}^{\rm HD} \left[\frac{\delta_{\ell\ell'}}{n_\ell}\E(H_u^2+2H_uH_{B}) + \Var(H_2)\right].
\end{equation}
The first term represents interference-induced angular fluctuations at fixed source powers, whereas the second arises from population-to-population fluctuations in the total GW power and correlates different multipoles.

Conditional narrowing therefore does not necessarily imply a smaller population variance. At fixed total power $H_2$, the interference-induced variance relative to the Gaussian case is suppressed by the factor $1-f_B^2$. By contrast, if $H_u$ is held fixed while a bright source is added, the absolute conditional variance is enhanced relative to the unresolved-background value by the factor
$1+2\eta_B$. A deterministic single source has no pulsar-averaged interference-induced angular fluctuations, although its response to a fixed finite pulsar array remains direction dependent.


For several explicit sources with fixed powers and with their phases and sky directions averaged over, define $H = H_u + \sum_j |s_j|^2 $ and $H_4 = \sum_j|s_j|^4,$, where the sums run over the explicit discrete sources. The normalized multipole variance then becomes
\begin{equation}
\Var(Q_\ell \mid \langle \bm b^B_\ell \rangle_{\Phi, \hat{\bm \Omega}}, H_u)=\frac{1-r}{n_\ell}, \qquad r = \frac{H_4}{H^2},
\end{equation}
where $\langle \rangle_{\Phi, \hat{\bm \Omega}}$ denotes the average over phases and sky directions.
For a single explicit source, $r=f_B^2$, recovering Eq.~\eqref{eq:fraction}. The Gaussian unresolved component contributes no discrete $|s_j|^4$ term in this hybrid model. Replacing a genuinely finite faint population by a Gaussian remainder therefore neglects its source-level connected higher-order cumulants.

\begin{table}[t]
\caption{
Schematic regimes of the bright-source and background model.
}
\begin{ruledtabular}
\begin{tabular}{lccc}
Regime & $N_{\rm eff}$ & $f_B$ & $S_{\ell }$ statistics \\
\hline
Many faint sources & $\gg1$ & $\ll1$ & Gamma \\
Bright + faint & $\gg1$ & $0<f_B<1$ & non-central $\chi^2$ \\
Bright dominated & --- & $\sim 1$ & CW-like limit
\end{tabular}
\end{ruledtabular}
\label{tab:regimes}
\end{table}

\section{Autocorrelations}

The pulsar autocorrelation requires separate treatment from the cross-correlation between distinct pulsars. Restoring the pulsar term and averaging over pulsar directions and the rapidly varying pulsar phase gives
\begin{equation}
\left\langle \tilde z_a\tilde z_a^{*} \right\rangle = \sum_{\ell=2}^{\infty}
S_\ell + \frac{H_2}{2},
\label{auto-correlation}
\end{equation}
in agreement with Ref.~\cite{Bi:2023tib}.
The additional term $H_2/2$ arises from the pulsar contribution to the autocorrelation. It modifies the diagonal pulsar covariance but is not part of the Legendre expansion describing the cross-correlation between distinct pulsars.

Within the many-source approximation and at fixed $H_2$, Eq.~\eqref{auto-correlation} is a shifted sum of independent Gamma or non-central-$\chi^2$ distributed multipoles with different scales. It therefore does not, in general, reduce to a single Gamma distribution. After marginalizing over the population-power distribution $p_H(H_2)$, the autocorrelation distribution can be evaluated numerically, for example by Monte Carlo sampling or numerical inversion of its Laplace transform.

\end{document}